\documentclass[12pt,a4paper,final]{iopart}

\usepackage{iopams}  
\usepackage{graphicx}
\usepackage[breaklinks=true,colorlinks=true,linkcolor=blue,urlcolor=blue,citecolor=blue]{hyperref}
\usepackage{color}
\usepackage{url}

\begin{document}

\title[Microtearing modes as the source of magnetic fluctuations in the JET pedestal]{Microtearing modes as the source of magnetic fluctuations in the JET pedestal}

\author{D. R. Hatch$^{1*}$,~M. Kotschenreuther$^{1}$,~S. M. Mahajan$^{1}$,~M. J. Pueschel$^{1}$,~C. Michoski$^{2}$,~G. Merlo$^{2}$,~E. Hassan$^1$,~A. R. Field$^{3}$,~L. Frassinetti$^{4}$,~C. Giroud$^{3}$,~J. C. Hillesheim$^{3}$,~C. F. Maggi$^{3}$,~C. Perez von Thun$^{5}$,~C. M. Roach$^{3}$,~S. Saarelma$^{3}$,~D. Jarema$^{6}$,~F. Jenko$^{1,2,6}$,~JET contributors$^{7}$}
\address{$^1$Institute for Fusion Studies, University of Texas at Austin, Austin, Texas, 78712}
\address{$^2$Oden Institute for Computational Engineering and Sciences, University of Texas at Austin, Austin, Texas, 78712}
\address{$^3$CCFE, Culham Science Center, Abingdon OX14 3DB, United Kingdom}
\address{$^4$Division of Fusion Plasma Physics, KTH Royal Institute of Technology, SE-10691 Stockholm, Sweden}
\address{$^5$Institute of Plasma Physics and Laser Microfusion, Hery 23, 01-497 Warsaw, Poland}
\address{$^6$Max-Planck-Institute for Plasma Physics, Boltzmannstrasse 2, 85748 Garching, Germany}
\address{$^7$See the author list of X. Litaudon et al 2017 Nucl. Fusion 57 102001}
\ead{$^*$drhatch@austin.utexas.edu}



\begin{abstract}

We report on a detailed study of magnetic fluctuations in the JET pedestal, employing basic theoretical considerations, gyrokinetic simulations, and experimental fluctuation data, to establish the physical basis for their origin, role, and distinctive characteristics.  We demonstrate quantitative agreement between gyrokinetic simulations of microtearing modes (MTMs) and two magnetic frequency bands with corresponding toroidal mode numbers n=4 and 8.  Such disparate fluctuation scales, with substantial gaps between toroidal mode numbers, are commonly observed in pedestal fluctuations.  Here we provide a clear explanation, namely the alignment of the relevant rational surfaces (and not others) with the peak in the $\omega_*$ profile, which is localized in the steep gradient region of the pedestal.  We demonstrate that a global treatment is required to capture this effect.  
Nonlinear simulations suggest that the MTM fluctuations produce experimentally-relevant transport levels and saturate by relaxing the background electron temperature gradient, slightly downshifting the fluctuation frequencies from the linear predictions.    
Scans in collisionality are compared with simple MTM dispersion relations.  At the experimental points considered, MTM growth rates can either increase or decrease with collision frequency depending on the parameters thus defying any simple characterization of collisionality dependence.  

\end{abstract}

\pacs{00.00}
\vspace{2pc}
\section{Introduction} \label{introduction}


High frequency magnetic fluctuations are often observed in the edge of H-mode discharges\cite{perez_03,diallo_14,diallo_15,laggner_16,cavedon_19,chen_19,laggner_19}.  On JET, these fluctuations have been dubbed washboard modes~\cite{perez_03} due to their distinctive appearance in magnetic spectrograms and in lieu of a clearly identified underlying physical mechanism.  These fluctuations are localized in or near the pedestal; arise during the inter-ELM cycle; are correlated with the saturation of pedestal temperature (in contrast with density); and have frequencies in the electron diamagnetic direction in the plasma frame~\cite{perez_03}.  In this paper, we study such fluctuations using basic theoretical considerations along with gyrokinetic simulations using the \textsc{Gene} code~\cite{jenko_00b,goerler_11} and unambiguously identify them as microtearing modes (MTMs), thus establishing an increasingly firm physical basis for their origin, role, and distinctive characteristics.

The microtearing mode was first described in Ref.~\cite{HDW}, which established already the central characteristics of the basic instability: its intrinsic electromagnetic nature, a reliance on mode frequency being roughly comparable to the collision frequency, its drive by the electron temperature gradient, and a frequency close to $\omega_{e*} = k_y \rho_s c_s (1/L_n + 1/L_{Te})$ (here, $k_y$ is the binormal wavenumbers, $\rho_s$ is the sound gyroradius, $c_s$ is the sound speed, and $L_{n,Te}$ is the gradient scale length of the density or temperature).  Subsequent work~\cite{mahajan_79,gladd_80} elucidated the role of the parallel wavenumber, $k_{||}$, the separate influence of density and temperature gradients, and the possibility of collisionless variations on the theme~\cite{carmody_13,predebon_13,swamy_14,hamed_18}.  Gyrokinetic simulations have further explored the nonlinear saturation physics~\cite{doerk_11,doerk_12,guttenfelder_11}, and its role in spherical tokamaks (both core~\cite{applegate_07,guttenfelder_11,chowdhury_16}, and edge~\cite{dickinson_12,canik_13}).  In an interesting recent development, Ref.~\cite{jian_19} identifies MTM as the salient electron heat transport mechanism in an internal transport barrier on DIII-D, thus establishing a unifying connection with the present work: the activity of MTM in transport barriers after the conventional turbulent transport mechanisms have been suppressed.

This paper builds on a series of recent papers that demonstrate that MTM is a prominent fluctuation in the steep gradient region of the pedestal~\cite{hatch_16, kotschenreuther_19}.  Ref.~\cite{hatch_16} identified MTM as the salient ion-scale instability in a JET ITER-like wall (JET-ILW) discharge, noted its unusual mode structure (peaking at the top and bottom of the tokamak), and demonstrated that nonlinear simulations can reproduce experimental transport levels.  Ref.~\cite{kotschenreuther_19} argued that MTMs are likely responsible for pedestal transport on many discharges based on their distinctive physical characteristics, placed MTMs among other prospective pedestal transport mechanisms, and demonstrated close correspondence between GK simulations and magnetic spectrograms for two DIII-D discharges.  Recently, experimental work has exploited an innovative measurement of internal magnetic fluctuations~\cite{chen_16} to further establish MTM as a common pedestal fluctuation in DIII-D~\cite{chen_19}.

This paper expands this line of research in several ways: (1) demonstrating quantitative agreement between between GK simulations and distinctive features of magnetic spectrograms, namely discrete bands of frequencies at disparate toroidal mode numbers, $n$ (Sec.~\ref{magnetic_fluctuations}); (2) providing a clear explanation for the discrete band structure (Sec.~\ref{bands}); (3) elucidating the underlying physics by comparing MTM collisionality dependence with a simple dispersion relation (Sec.~\ref{sec:dispersion}); (4) demonstrating that global simulations are necessary to quantitatively reproduce the experimental fluctuations (Sec.~\ref{local_v_global}); (5) elucidating a possible nonlinear saturation mechanism and demonstrating experimentally realistic transport levels (Sec.~\ref{sec:nonlinear}); and (6) discussing the role of MTM in limiting the pedestal and influencing inter-ELM pedestal evolution (Sec.~\ref{interELM}). 

\section{Magnetic Fluctuations in JET pulse 78697} \label{magnetic_fluctuations}

\begin{figure}[htb!]
 \centering
 \includegraphics{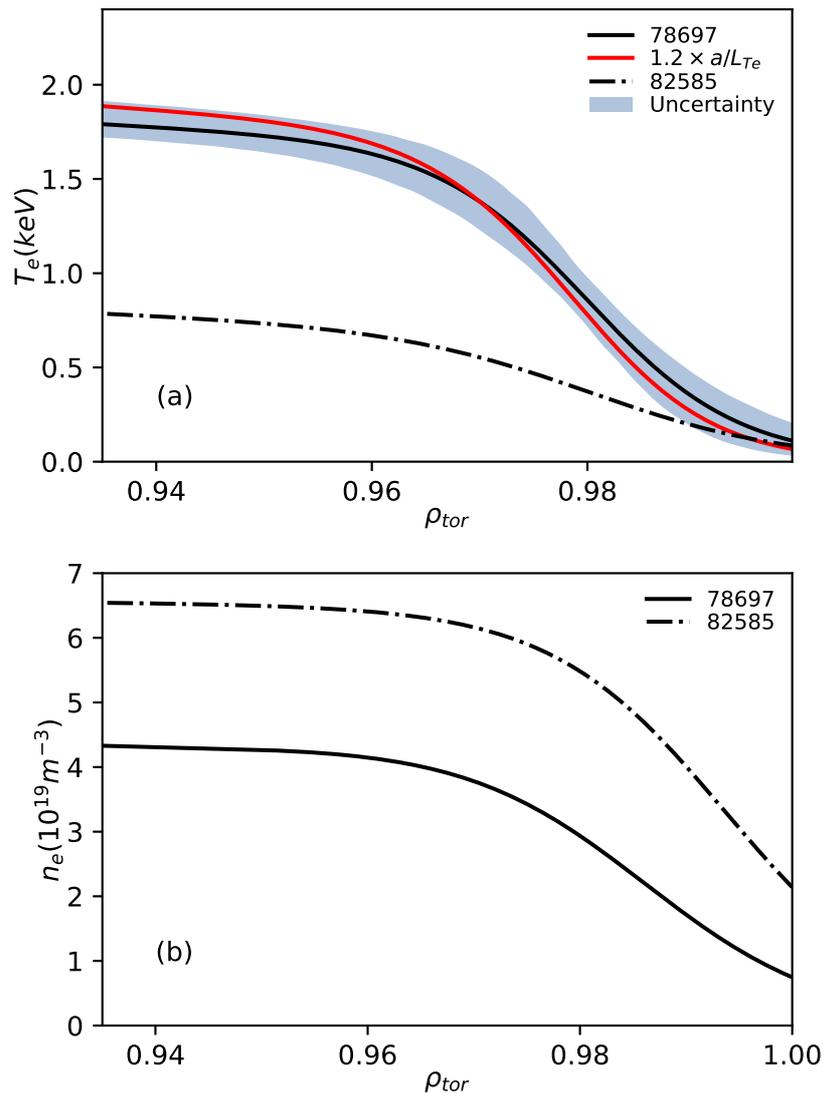}
 \caption{\label{profiles_w_uncertainty} Profiles of electron temperature (top) and electron density (bottom) for three JET discharges.  The shaded band around the 78697 $T_e$ profile signifies the $\pm 1 \sigma$ uncertainty in the profile and the red line is an additional $T_e$ profile (within the uncertainty band) used for some simulations.   }
\end{figure}

JET pulse 78697 is a high current, high field (3 MA, 2.4 T) carbon wall discharge characterized by parameters shown in Table 1 and profiles shown in Fig.~\ref{profiles_w_uncertainty}.  It was examined in detail in a comparative study of pedestal transport on JET~\cite{hatch_19}, for which it was selected as a carbon wall counterpart to a high-performance JET discharge (pulse 92432) from the ITER-like wall (ILW) period.  MTMs were also studied in an additional JET discharge (pulse 82585) as reported in Ref.~\cite{hatch_16}.  This paper focuses mostly on pulse 78697, for which we have carried out a detailed comparison with magnetic fluctuation measurements.  Some results from 82585 will also be discussed.  Pedestal profiles of electron temperature and density are shown in Fig.~\ref{profiles_w_uncertainty}.  The uncertainty band shown in Fig.~\ref{profiles_w_uncertainty} is calculated using a simple Monte Carlo method.  One thousand modified tanh functions were created with different input parameters.  The input parameters are determined from the the average plus a random value sampled from a Gaussian distribution centered at 0 and with one standard deviation uncertainty in the parameters.  The upper and lower bounds in the figure highlight the region within one standard deviation.  

\begin{figure}[htb!]
 \centering
 \includegraphics[angle=90,scale=0.7]{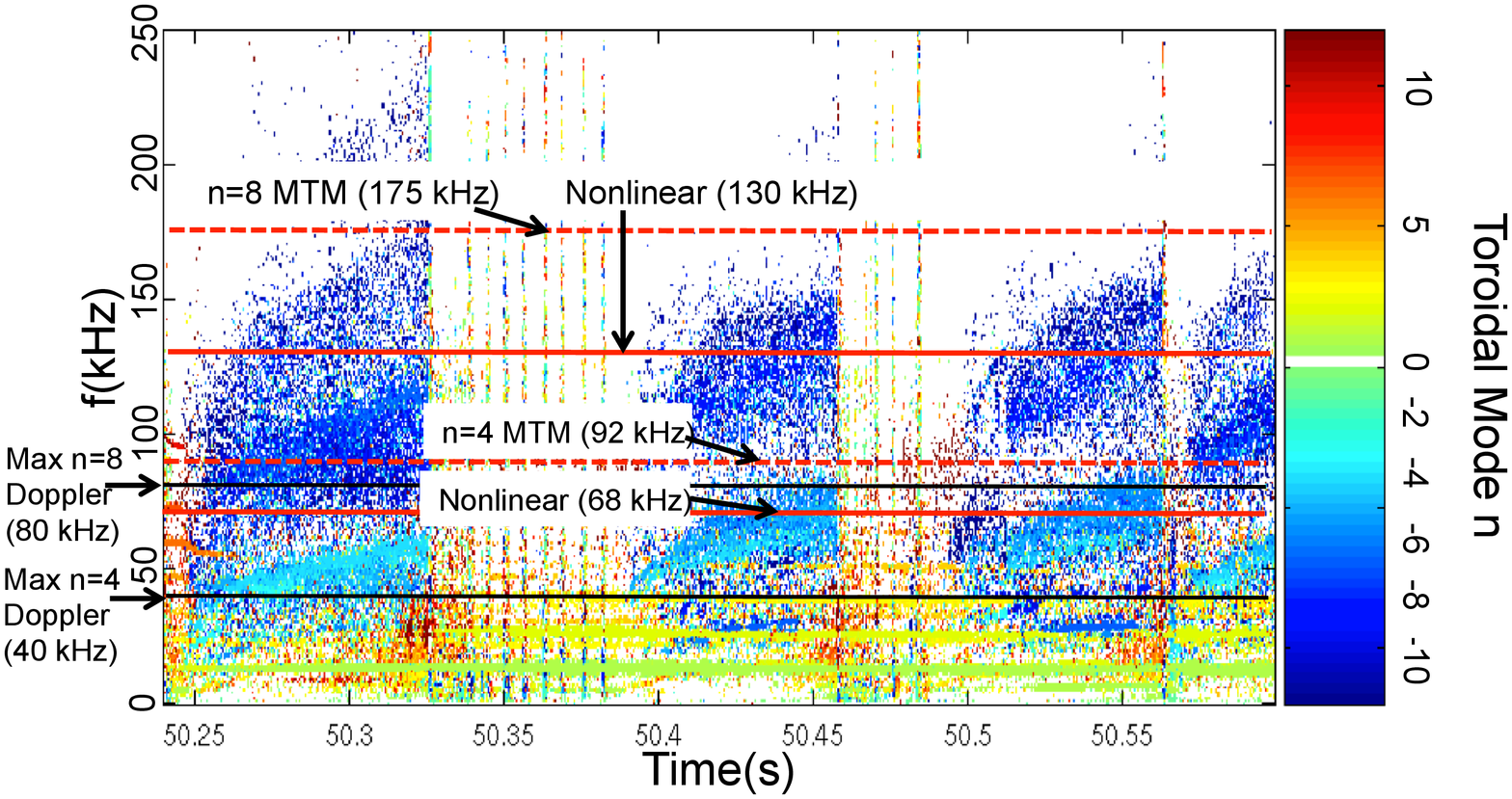}
 \caption{\label{washboard_plot}  Magnetic spectrogram for JET-C (78697) showing frequency bands and toroidal mode numbers of magnetic fluctuations.  The color denotes the toroidal mode number, and the sign of the toroidal mode number denotes the propagation direction (negative---electron diamagnetic).  The blue high frequency bands correspond to washboard modes and are the focus of this paper.  We focus, in particular on the two central inter-ELM periods ($t(s)=50.4-50.55$)  The abrupt cessations of the washboard fluctuations correspond to ELM crashes. The green lower frequency bands denote core modes, which are not correlated with the inter-ELM behavior and not of interest in this paper.  The dashed lines denote the frequencies of MTMs from global gyrokinetic simulations carried out in the lab frame ($n=8$ at $\sim174 kHz$, and $n=4$ at $92 kHz$).  The lower solid red lines show frequencies from nonlinear simulations.  The maximum Doppler shifts (black) from $E_r \times B$ rotation are shown in black, demonstrating that the washboard modes are in the electron direction in the plasma frame. }
\end{figure}

\subsection{Magnetic Fluctuation Data} \label{fluctuation_data}

Mirnov coils on JET often identify magnetic fluctuations, so-called washboard modes, with distinct frequency bands~\cite{perez_03}.  Ref.~\cite{perez_03} determined that these fluctuations are localized in or near the pedestal; arise during the inter-ELM cycle; are correlated with the saturation of pedestal temperature (in contrast with density); and have frequencies in the electron diamagnetic direction in the plasma frame.  All these properties are consistent with the MTM, which is reinforced by the simulation results described below.  

Fig.~\ref{washboard_plot} shows the magnetic spectrogram for JET pulse 78697.  The color denotes the toroidal mode numbers extracted from the phase shift of the signal from a toroidal set of Mirnov coils (the sign indicates electron [negative] and ion [positive] diamagnetic direction).  The low frequency (green) bands are core modes that are not correlated with the inter-ELM cycle and not of interest for the present study.  We are interested, rather, in the high frequency bands propagating in the electron diamagnetic direction (blue). Four inter-ELM cycles are captured in Fig.~\ref{washboard_plot}.    Here, we focus on the two middle ELM cycles in the time range $t \sim 50.4s - 50.55s$, which exhibit slightly different frequency bands from the first and last inter-ELM cycles.    At least two distinct bands are manifest:  a lower frequency band ($f \approx 60-80 kHz$) with dominant contributions from $n \sim -4,-5$ and a higher frequency band ($f \sim 100-140 kHz$) with dominant contributions from $n \approx -7,-8,-9$.  This can be viewed, roughly, as two bands at $n=-4 \pm 1$ and $n=-8 \pm 1$.  As will be described below, gyrokinetic simulations produce MTMs with these characteristics.  

In order to compare simulations with these lab-frame measurements, the rotation must be carefully accounted for.  Direct measurements of the radial electric field $E_r$ and parallel flow $V_{||}$ were not available for this discharge.  Consequently, we estimate $E_r$ using the standard neoclassical formula~\cite{landreman_12},
\begin{equation}
V_{||} = -\frac{R B_{\phi}}{ZeB} \left (\frac{1}{n_i}\frac{dP_i}{d\psi} + Ze\frac{d\Phi_0}{d\psi} -K\frac{B^2}{<B>^2}\frac{dT_i}{d\psi}\right ).
\label{Er}
\end{equation}
where $V_{||}$ is the parallel flow, $R$ is the major radius, $B_{\phi}$ is the poloidal magnetic field, $P_i$ is the ion pressure, $\Phi_0$ is the electrostatic potential, and $\psi$ is the normalized poloidal magnetic flux (readers are referred to Ref.~\cite{landreman_12} for more detailed definitions, e.g., of $K$).  Since there is no measurement available, we use the approximation $V_{||}=0$, which is justified by the fact that in other pedestal scenarios where measurements are available, the dominant balance in Eq.~\ref{Er} is between the radial electric field and the gradients~\cite{viezzer_13}.  Note that the Doppler shift in the pedestal is in the electron diamagnetic direction and thus opposite to that of the bulk plasma rotation in the core, so the inclusion of $V_{||}$ would be expected to slightly decrease the net Doppler shift.  
The resulting $E \times B$ advection from the radial electric field is included in the simulations described below so that the simulations are carried out in the lab frame.  The resulting Doppler shift is $\Omega = \frac{n E_r}{R B_\theta}$.   

The maximum (over radial location) Doppler shifted frequency is a useful discriminator between ion frequency fluctuations, like kinetic ballooning modes (KBMs), and fluctuations with frequencies strongly in the electron direction like MTMs.  These maximum Doppler frequencies, shown in Fig.~\ref{washboard_plot} for $n=4,8$ lie at the lowest measured frequency of the washboard fluctuations in the first inter-ELM cycle and well below the measured frequency for the other ELM cycles.  This is a clear indication that the fluctuations are propagating in the electron diamagnetic direction in the plasma frame.  


We also note in passing an additional narrower band with $n=-7$, which is observed in the second and third ELM cycles with $f \sim 30 kHz$.  This frequency is far below the corresponding $n=7$ Doppler shift.  Possible explanations include (1) a MTM localized at the pedestal top where both the Doppler shift and $\omega_{e*}$ are much smaller (note that pedestal-top MTMs have been described in Ref.~\cite{dickinson_12,dickinson_13,saarelma_13,chowdhury_16}), and (2) a KBM or ITG mode with ion frequency in the plasma frame.  

\section{Comparison between Linear Simulations and Spectrograms} \label{comparison}

Global linear simulations for JET discharge 78697 were performed with minor modifications to the simulation inputs to test sensitivities.  Numerical details are described in Appendix A.  Two cases will be discussed in detail: (1) the base case comprising the nominal profiles and equilibrium reconstruction, and (2) a scenario (labeled `Mod 1') with a $5 \%$ reduction in safety factor $q$ and a 20\% increase in the electron temperature gradient (maintaining constant total pressure by decreasing the ion temperature gradient correspondingly).  The relevance of the modified $q$ profile will be discussed below.  The $20 \%$ increase in electron temperature gradient is within the error bars of the experimental measurement as shown in Fig.~\ref{profiles_w_uncertainty}.  Likewise, the minor variation of the $q$ profile is also likely within experimental uncertainty.  To demonstrate this, we reconstructed a self-consistent equilibrium prescribing a similar q profile and allowing the equilibrium reconstruction to adapt.  Resulting simulations are very similar to those described below. 

\begin{figure}[htb!]
 \centering
 \includegraphics{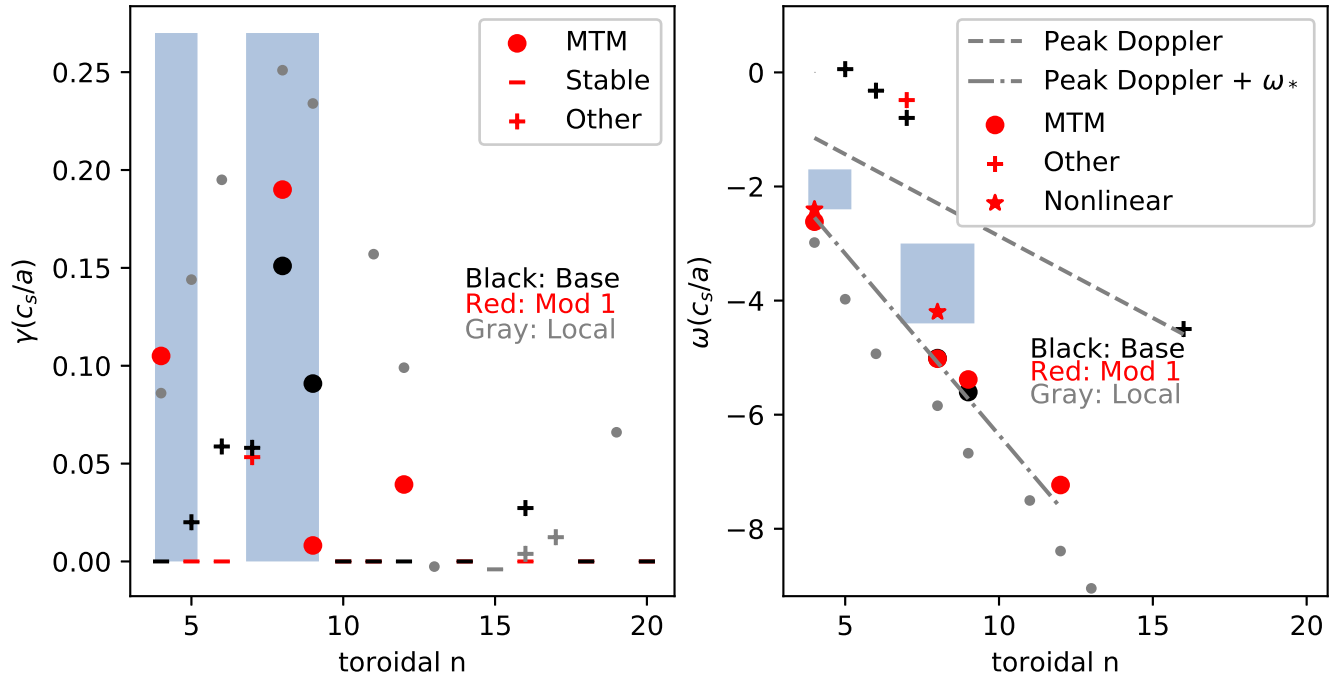}
 \caption{\label{gamma_ky} Growth rates and frequencies from gyrokinetic simulations for two cases.  The base case uses the nominal experimental inputs and the Mod 1 case uses a slightly shifted q profile and a $20 \%$ increase in the electron temperature gradient (shown in Fig.~\ref{profiles_w_uncertainty}).  The circles denote MTMs and the plus symbols denote other modes. The blue shaded regions signify the experimental values from the magnetic spectrogram.  The red stars denote the frequencies calculated in a nonlinear simulation.  The MTMs from the Mod 1 case agree quantitatively with the toroidal mode numbers and frequencies observed in the spectrogram.    }
\end{figure}

Fig.~\ref{gamma_ky} shows growth rates (left) and frequencies (right) for these two cases: the Base case in black and Mod 1 in red.  MTMs are denoted with circles and other modes with plus symbols.  For the base case, MTMs are observed at $n=8,9$ (in agreement with the spectrogram) but not at $n=4$.  The instability that is found at $n=4$ does not correspond to the observed fluctations, having a much lower frequency.  For the Mod 1 case, MTMs are unstable at $n=4,8,9$, in quantitative agreement with the experimental observation, whose range in the spectrogram is denoted by the shaded blue regions.  Likewise, there is reasonable agreement between simulated and observed frequencies, with the simulated frequencies surpassing the observed frequencies by $\sim 30\%$.  Quantitative agreement is achieved in nonlinear simulations, which will be described in Sec.~\ref{nonlinear}.  Notably, the most distinctive feature of the spectrograms---the quasi-coherent bands at disparate $n$ numbers---is reproduced by the simulations in the sense that unstable MTMs are observed at toroidal mode numbers $n=4, 8, 9$ but not those in between ($n=5,6,7$).  The underlying mechanism for this distinctive behavior is elucidated in the next subsection.  

For reference, results from local MTM simulations for the Mod 1 case located at the peak in the global mode structure---$\rho_{tor} = 0.978$---are shown in Fig.~\ref{gamma_ky} in gray.  They exhibit similar growth rates to the global simulations only at mode numbers where MTMs are strongly unstable in the global simulations.  Notably, inconsistent with the experimental observations, the local growth rates also exhibit instability at the intermediate $n$ numbers.  Local simulations are discussed further in Sec.~\ref{local_v_global}, where it is concluded that the local approach is inadequate for capturing the distinctive features of pedestal MTMs and, moreover, can be quite misleading. 

\section{Explanation for MTM frequency bands } \label{bands}

\begin{figure}[htb!]
 \centering
 \includegraphics[scale=0.7]{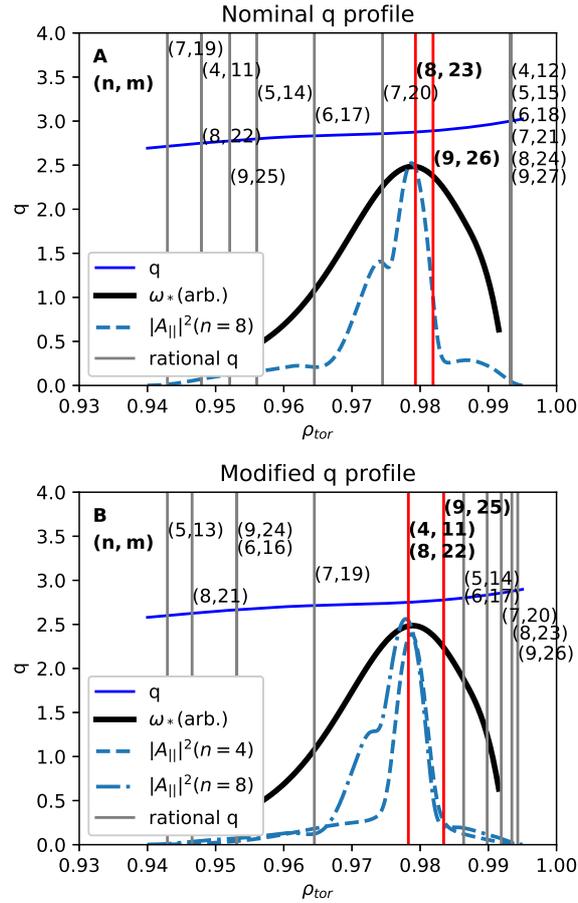}
 \caption{\label{oms_q_Apar} q profiles, $\omega_*$ profiles, and MTM mode structures for the Base and Mod 1 cases.  The vertical lines correspond to rational surfaces and are labeled by their corresponding $n$ and $m$ numbers.  For the Base case, simulations find unstable MTMs only at $n=8,9$.  For the Mod 1 case $n=4,8,9$ are found to be unstable in agreement with the experimental observations.  MTMs are unstable only when the appropriate rational surfaces align with the peak in the $\omega_*$ profile, which offers a clear explanation for the bands observed in the spectrogram.  }
\end{figure}

Perhaps the most important result of this work is a clear explanation for the distinct fluctuation bands at disparate n numbers.  The key insight follows from a careful examination of the alignment of the MTM drive, i.e. $\omega_{e*} = k_y \rho_s c_s (1/L_n + 1/L_{Te})$  with the relevant rational $q$ surfaces, as shown in Fig.~\ref{oms_q_Apar}.  In this figure, the $\omega_{e*}$ profile (black) is shown along with the $q$ profile (blue) and representative linear eigenmode structures (dashed blue).  Vertical lines denote rational surfaces, labeled with their respective toroidal and poloidal mode numbers.  Fig.~\ref{oms_q_Apar} A. shows the base case, where $(n,m)=(8,23)$ and $(n,m)=(9,26)$ are the surfaces that align most closely with the peak in the $\omega_{e*}$ profile, corresponding to the unstable MTM at $n=8,9$ (recall Fig.~\ref{gamma_ky}).  Notably, the $n=4$ rational surfaces lie far from the peak in $\omega_{e*}$ and there is no $n=4$ instability that corresponds to the fluctuation data.  In contrast, Fig.~\ref{oms_q_Apar} B. shows results for the Mod 1 case (differing in a $\sim 5\%$ decrease in the $q$ profile).   In this case, the $(n,m)=(4,11)$ and $(n,m)=8,22)$ rational surfaces now both align with the peak in $\omega_{e*}$, resulting in unstable mode numbers that correspond quantitatively with the magnetic spectrogram.  We thus have a clear physical explanation for the discrete frequency bands observed in the data, namely the alignment of particular rational surfaces with the peak in the $\omega_*$ profile.  


\subsection{Local vs. Global} \label{local_v_global}

\begin{figure}[htb!]
 \centering
 \includegraphics[scale=0.8]{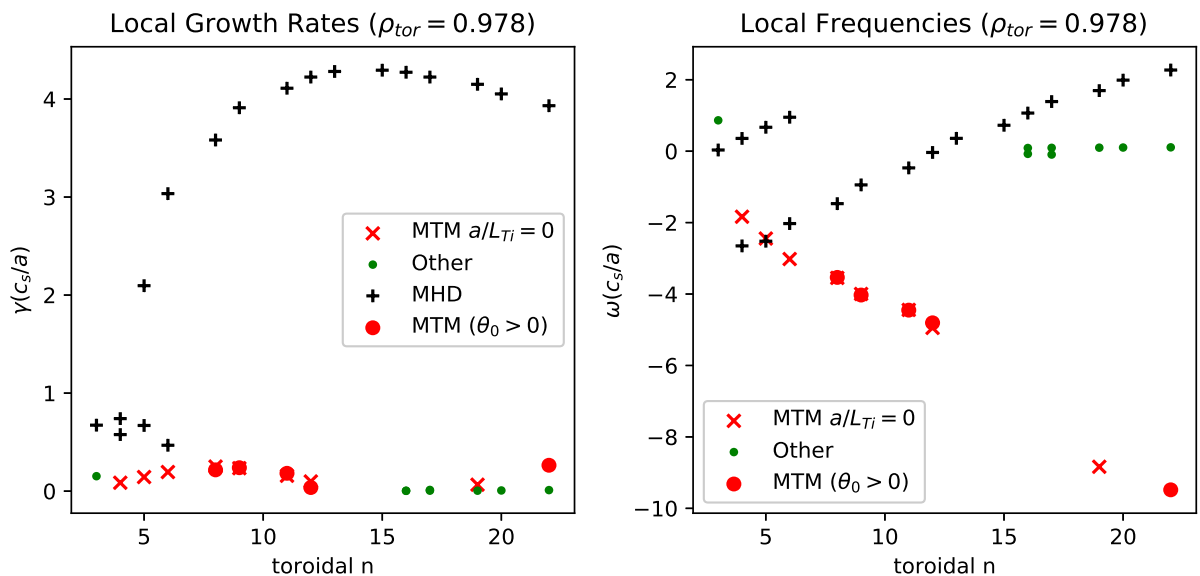}
 \caption{\label{gamma_ky_local}  Growth rates and frequencies from local linear simulations.       }
\end{figure}

Local simulations were carried out centered at the radial location where the global modes peak ($\rho_{tor} = 0.978$).  As described in Ref.~\cite{hatch_16}, MTMs can be unstable at finite $\theta_0$, requiring a scan of this dimension in the local simulations.  Growth rates and frequencies are shown in Fig.~\ref{gamma_ky_local}.  A strongly unstable MHD-like mode exists at this radial location as denoted with the black symbols.  The frequencies for these modes are unconventional in that they are negative at low $n$ and transition to positive as $n$ increase. The mode structure of the low $n$ modes, together with the fact that the mode disappears when ion temperature gradients are eliminated, is suggestive of past descriptions of ideal MHD modes that are destabilized below the ideal threshold by kinetic effects arising at large ballooning angle. The tail of this mode has a long decaying envelope in theta, with oscillations of $2 \pi$, similar to these. The modes in the early literature~\cite{hastie_81,cheng_82,kotschenreuther_86} have positive mode frequency (ion direction) but are also described in circular geometry, and it is possible that strongly shaped geometry (i.e., near the separtrix) could lead to a negative frequency. These modes are more difficult to destabilize when the ideal MHD mode is further below marginal stability. This might also explain why these modes disappear in global calculations---the ideal ballooning mode is more stable in such global calculations as described in Refs.~\cite{hatch_16,kotschenreuther_19,hatch_19}.   This phenomenon was explained by noting that the effective radial extent of the local modes (dominated by $k_x=0$) is inconsistent with the physical extent of its instability drive, which is limited to the steep gradient region of the pedestal (readers are referred to Refs.~\cite{kotschenreuther_19,hatch_19} for further discussion and a criterion for when this phenomenon may be expected to occur).  In addition  to their stability in the global simulations, the experimental data also is inconsistent with the characteristics of these modes: (1) the frequency of the $n=4$ mode is roughly in the right range, but the trend in the dispersion relation is opposite that of the experimental observation, and (2) the modes are unstable at each consecutive $n$ as opposed to the distinct and separated $n$ numbers observed in global simulations of MTMs.  We conclude that, for the purpose of explaining the experimental characteristics, the focus on MTMs in the previous sections is justified and a global treatment is likely necessary to quantitatively reproduce the observations.  

The subdominant MTM instabilities can be identified either by considering higher $\theta_0$ (denoted by red dots in Fig.~\ref{gamma_ky_local}) or by eliminating the ion temperature gradient (red x symbols), which stabilizes the fastest growing mode while leaving the MTMs unchanged.  As noted above in the discussion of Fig.~\ref{gamma_ky}, the local MTMs have roughly the same growth rates as the global calculations for $n=4,8$.  However, in contrast with global, the local MTMs exhibit a smooth dispersion relation, with instabilities at each consecutive $n$.  This is easily explained by noting that (1) in a local treatment, $\omega_*$ is defined by its local value and (2) the $q$ profile is linear (i.e. has constant $\hat{s}$).  Consequently, different $n$ numbers whose rational surfaces may be slightly offset from each other still sample the same value of $\omega_*$, in contrast with the global treatment (recall Fig.~\ref{oms_q_Apar}). 

We conclude that the local approximation is inadequate for quantitatively reproducing the observed magnetic flucatuations and, moreover, may produce artificially unstable MHD modes.  It remains to be seen whether alternative approaches to capturing nonlocal effects~\cite{parra_15,abdoul_15} would be sufficient.  The present scenario would be a compelling test of such capabilities.

\Table{\label{tab1} Summary of important parameters for JET pulse 78697 (C).  $I_p$ is the plasma current, $B_T$ is the toroidal magnetic field, $q_{95}$ is the safety factor at $95\%$ of the minor radius (in terms of normalized poloidal magnetic flux), $\delta$ is triangularity, $P_h$ is the total heating power, $P_i$ is the inter-ELM power loss~\cite{field_20}, Gas is the fueling rate, $\beta_N$ is normalized plasma pressure, $T_{e,p}$ is the pedestal top electron temperature, and $n_{e,p}$ is the pedestal top electron density.  }
\br
\centering
Pulse & $I_p(MA)$ & $B_T(T)$ & $q_{95}$ & $\delta$ & $P_{h} (MW)$ & $P_{i} (MW)$ & Gas(e/s) & $\beta_N$ & $T_{e,p}(keV)$ & $n_{e,p} (10^{19}m^{-3})$  \\
\mr
78697 & 3.0  & 2.4 & 2.6 & 0.24 & 14.8 & 5.7 & 0.0 &1.8 & 1.68 & 4.19   \\
\br
\end{tabular}
\end{indented}
\end{table}

\subsection{Simple MTM Dispersion Relation} \label{sec:dispersion}

\begin{figure}[htb!]
 \centering
 \includegraphics{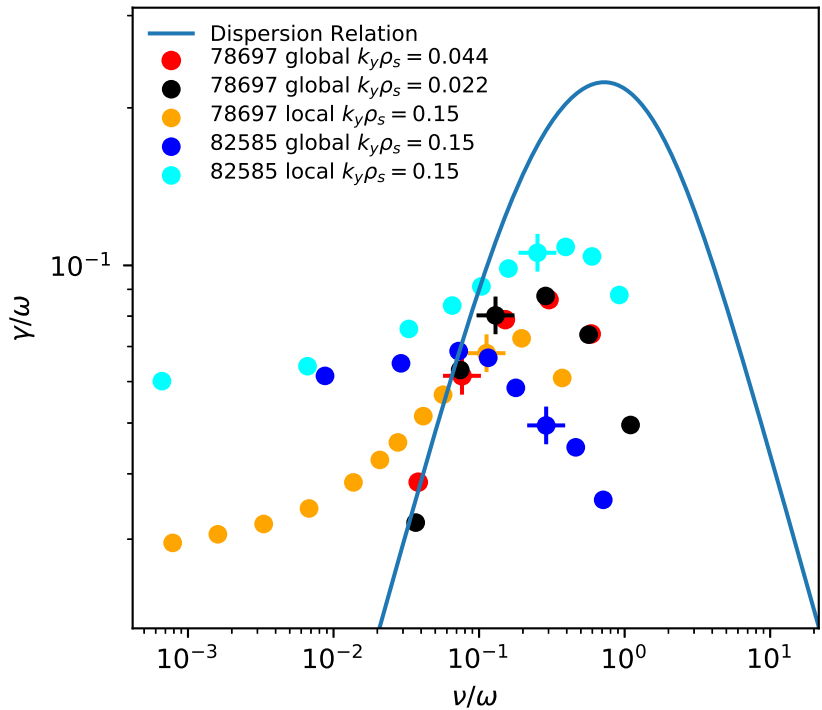}
 \caption{\label{MTM_dispersion} Comparisons between collisionality scans for various $k_y$ modes in two JET discharges and a simple MTM dispersion relation.  The simple dispersion relation qualitatively captures some of the collisionality dependence of the MTM.  In three of the cases, a collisionless tail is observed.  However, the experimental points (plus symbols) all correspond to a collisional or semi-collisional regime.     }
\end{figure}

In order to elucidate the underlying physics and further establish the modes of interest as MTMs, we make comparisons with the simple MTM dispersion relation from Ref.~\cite{HDW}

\begin{equation}
\left ( \nu - 0.54 i \omega \right ) \left ( \omega - \omega_{e*} \right ) - 0.8 \omega_{T*} \nu = 0 ,
\label{dispersion}
\end{equation}
where $\nu$ is the electron collision frequency, $\omega_{e*} = k_y \rho_s c_s (1/L_n + 1/L_{Te})$, and $\omega_{T*} = k_y \rho_s c_s (1/L_{Te})$.  More sophisticated versions are currently being studied and comparisons will be published elsewhere.  Fig.~\ref{MTM_dispersion} shows the solution to this dispersion relation along with simulated collisionality scans for both 78697 and 82585 for various $n$ numbers using both local and global modes of operation.  We focus first on the black and red symbols corresponding to the $n=4,8$ modes, respectively, for JET pulse 78697.  Although the simple dispersion relation differs substantially from the simulation results in the magnitude of the growth rate, it quantitatively captures the collisionality threshold and qualitatively captures other major features of the collisionality dependence, including a peak in growth rate at $\nu \sim \omega$ and subsequent falloff of growth rates at higher frequency.  This demonstration, in combination with other features of the mode, including predominantly electromagnetic heat flux, a large component of tearing parity, and $\omega = \omega_{e*}$, unambiguously identifies the modes as MTMs whose underlying physics connects directly with the earliest literature~\cite{HDW}.  Investigations of more comprehensive dispersion relations, which include $\eta$ (ratio of density gradient scale length to temperature gradient scale length) effects, finite $k_{||}$ effects, and a more sophisticated conductivity~\cite{larakers_20} are currently being undertaken and will be published elsewhere.

We briefly comment on the collisionality scans for 82585, which exhibit a tail in the collisionless limit.  We speculate that these modes retain some instability drive through a toroidal (curvature) or perhaps Alfv\'{e}nic resonance.  Such variations in MTM physics may be responsible for the multiple bands that are observed in some discharges.  For example, DIII-D discharge 162940 exhibits three distinct frequency bands: narrow bands at 40 kHz, 80 kHz, and a broader band centered at 400 kHz, which are in quantitative agreement with simulations as described in Ref.~\cite{hassan_20}.

We also note the plus symbols in Fig.~\ref{MTM_dispersion}, which correspond to the experimental values of collisionality.  The experimental points lie firmly in the range where the collisional effects are active.  In most of these cases, the experimental points lie below the peak of the dispersion relation, indicating that generally higher collisionality will destabilize the mode.  However, for the global simulation of discharge 82585, collisionality is clearly stabilizing, suggesting that the collisionality dependence may not be so easily characterized. Similarly to Fig. 6, Table 2 of Ref.~\cite{kotschenreuther_19} showed that the frequency of observed magnetic fluctuations of several discharges on ASDEX, DIIID and JET was consistent with MTM destabilization as in Eq.~\ref{dispersion}, suggesting that the physics simulated here applies to a wide class of pedestals.  

\section{Nonlinear Simulations} \label{sec:nonlinear}

Although the global nonlinear simulations described here offer several physical insights, they should be considered to be a qualitative representation of the dynamics due to certain numerical limitations.  The two main limitations are (1) numerical instabilities (possibly associated with low-n modes near ideal MHD limits) that preclude long-time simulations and extensive convergence tests, and (2) the artifical nature of the sources used to maintain the background profiles in gradient-driven simulations.  Solutions to both of these limitations are actively being developed by the \textsc{Gene} development team.  
Despite these limitations, several valuable insights can be gained.  

\begin{figure}[htb!]
 \centering
 \includegraphics{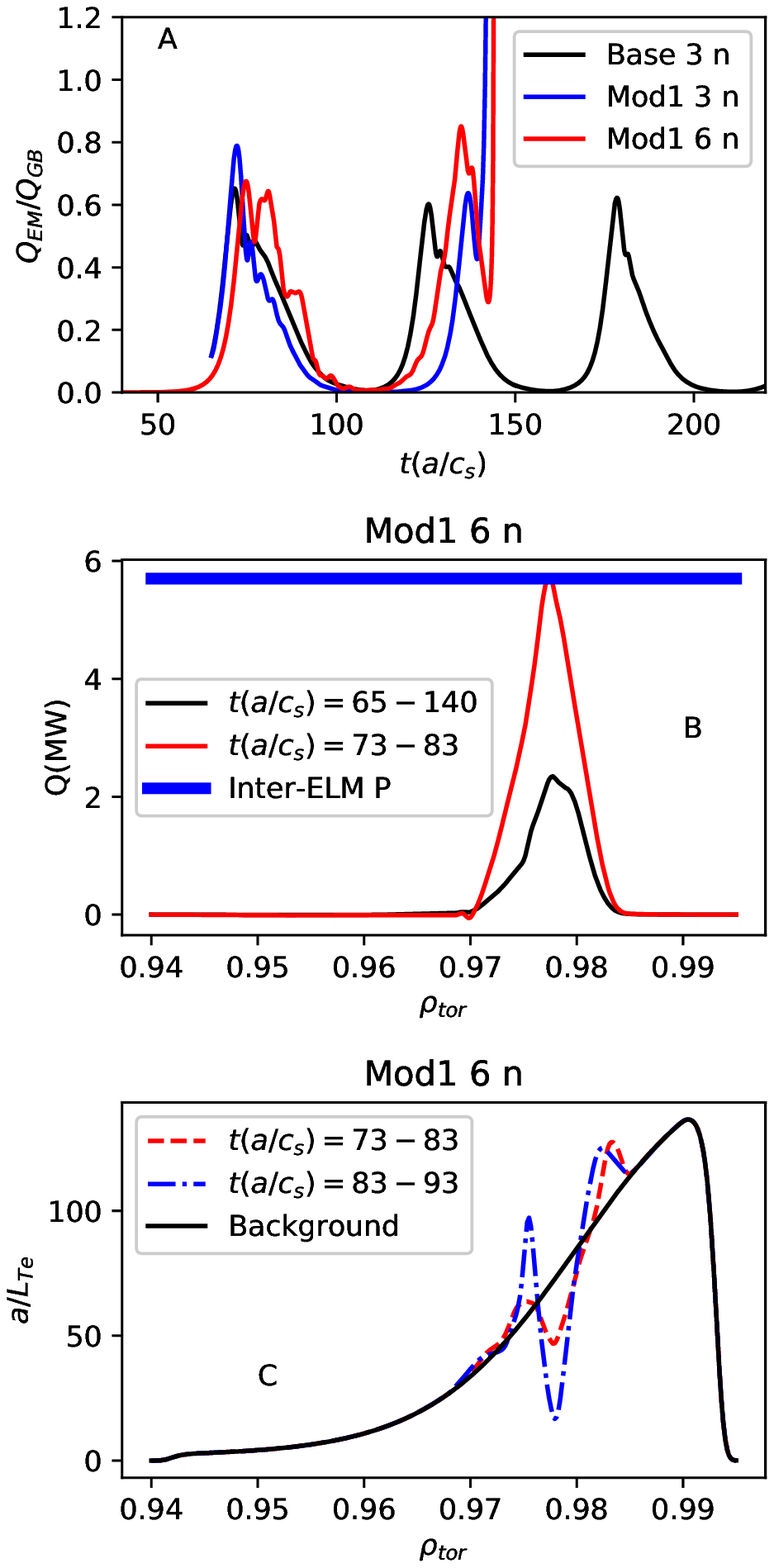}
 \label{nonlinear}
 \caption{\label{nonlinear} Time traces of electromagnetic heat flux (A.) for three simulations: the Base case with three toroidal mode numbers ($n=0,4,8$), the Mod 1 case with three toroidal mode numbers ($n=0,4,8$), and the Mod 1 case with six toroidal mode numbers ($n=0,2,4,6,8,10$).  This simulation with six modes is used for plots B. and C.  The latter two simulations end in numerical instabilities at $t(a/c_s) \sim 145$.  The radial profile of the electromagnetic heat flux (B.) averaged over two different time periods.  The radial profile of $a/L_{Te}$, demonstrating that the mode saturates by relaxing the background temperature gradient (C.).   }
\end{figure}

\label{Numerical Setup}

In order to simplify the dynamics, an adiabatic ion assumption is used, which results in minor quantitative changes for MTM stability.  In light of the above discussion, global simulations are necessary to capture the instabilities of interest.  As a first step, we retained only three toroidal mode numbers, $n=0$ along with the two mode numbers that appear most-prominently in the magnetic spectrogram: $n=4,8$.  Time traces for the electromagnetic heat flux are shown in Fig.~\ref{nonlinear} A. for the base case simulation (black) and the Mod 1 case (blue).  As a convergence test, six toroidal mode numbers are used for the Mod 1 case, ($n=0,2,4,6,8,10$), exhibiting only minor quantitative differences with the three mode simulation.  This simulation (six toroidal mode numbers) is the focus of the main results described below.  Background $E \times B$ shear is included in all nonlinear simulations.  Additional numerical details, including details on hyperdiffusion and source rates, are reported in Appendix A.    

In gradient-driven global simulations, source terms (Krook operators with time-independent coefficients) are employed in order to maintain the density and temperature profiles at their background values.  For these simulations, some transport characteristics are quite sensitive to the amplitude of these source terms.  When the sources are relatively weak, the background electron temperature profile is locally relaxed by the MTMs, thereby stabilizing the mode.  As the modes are stabilized due to the relaxed background gradients, the sources are then capable of restoring the background profiles at a rate dependent on the strength of the source coefficients and the instability grows once again.  This results in the cyclical behavior shown in Fig.~\ref{nonlinear} A.  If the sources are increased beyond a certain threshold, the decay does not occur, but the simulations end in numerical instability after a very brief saturation period.  Even when the sources are low, the simulations often end in numerical instability, as is seen in Fig.~\ref{nonlinear} A. for both the blue and red time traces.  Despite the sensitivity of the time traces to the source terms, other features of the simulations appear to be quite resilient to the details of the sources.  Notably, the peak of the heat flux is rather insensitive to the value of the sources, and is at a level that is quite close to the inter-ELM power loss, as seen in Fig.~\ref{nonlinear} B., where the radial profile of the electromagnetic heat flux is shown averaged over different time periods: the period of peak heat flux (red) and a longer time period that includes the decay phase (black).  

The nonlinear modification to the background temperature gradient $a/L_{Te}$ is shown in Fig.~\ref{nonlinear} C. for two time periods, showing a strong localized flattening of the temperature profile along with increases in the gradient on either side of the island region.  Assuming a statistical quasi-steady state, the outward radial heat flux from all transport mechanisms should be roughly constant across the domain.  Simulations of electron temperature gradient (ETG) transport for this discharge are reported in Ref.~\cite{hatch_19} and produced transport levels of $1-4 MW$ in the pedestal depending on radial location, suggesting it as the most likely mechanism responsible for the remaining electron heat transport.  For reference, the ETG simulation located at $\rho_{tor}=0.975$ is closest to the MTM location ($\rho_{tor}=0.978$) and produces 4 MW of electron heat transport.  In the actual physical system, MTMs and ETG are likely to vigorously interact, with, for example, MTMs working to flatten the temperature profile and ETG strongly stimulated in the steep gradient regions on either side of the island (see Fig.~\ref{nonlinear} C.) acting to reinforce the temperature gradient in the flattened region.  See Refs.~\cite{pueschel_20,maeyama_17} for additional work on multiscale interaction between MTMs and ETG.   

\begin{figure}[htb!]
 \centering
 \includegraphics{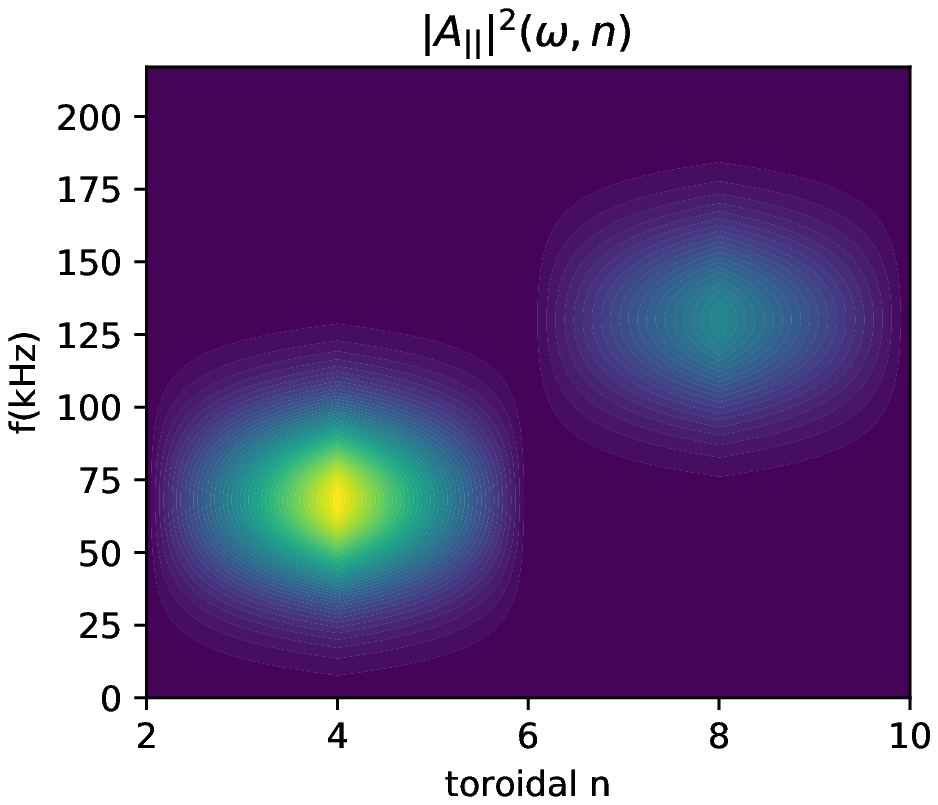}
 \caption{\label{frequency_spectrum} The frequency spectrum calculated during $t(a/c_s) = 73-83$ of the six mode nonlinear simulation.  This spectrum agrees closely with the $n$ numbers and frequency bands in the magnetic spectrogram (Fig.~\ref{washboard_plot}).   }
\end{figure}

The frequency spectrum from the nonlinear simulation is shown in Fig.~\ref{frequency_spectrum} resolved in toroidal mode number.  The precise values of the nonlinear frequencies are dependent on which phase of the nonlinear simulation is analyzed since the MTM frequency is very close to $\omega_*$, which decreases as the profiles relax.  The frequency spectrum shown in the figure is calculated over the time period of peak heat flux $t(a/c_s)=73-83$ and produces frequencies that are quite close to the experimental bands, albeit at the high range (see the red lines in Fig.~\ref{washboard_plot} and red stars in Fig.~\ref{gamma_ky}).  The frequency downshift is due to the relaxed background electron temperature gradient (i.e. a reduction in $\omega_{e*}$) that develops over the course of the simulation (recall Fig.~\ref{nonlinear} C.).  Averaging over the time period $t(a/c_s)=83-93$ results in an even larger reduction in the nonlinear frequencies (not shown).   

We emphasize the remarkable agreement between simulation and experiment exhibited in Fig.~\ref{frequency_spectrum}.  With modest variations of input parameters within experimental uncertainties, the simulations capture multiple features of the experimentally observed magnetic fluctuations, including the width and peaks of the magnetic fluctuation frequencies along with their corresponding toroidal mode numbers.      

\section{Discussion of inter-ELM Pedestal Dynamics} \label{interELM}

In the context of the results described above, we propose a picture of the role of MTMs in the inter-ELM evolution of the pedestal.  Following an ELM crash, the temperature and density profiles rebuild at different rates~\cite{wolfrum_09,diallo_15,laggner_16,maggi_17} and likely mediated by different transport mechanisms~\cite{kotschenreuther_19,TPT}.  In the context of MTM activity, we are particularly interested in the rebuilding of the electron temperature profile, which will steepen until the MTM surpasses some threshold.  The stability threshold will be crossed first at the radial location of peak drive (i.e. peak in $\omega_*$) and at toroidal mode numbers where two criteria are satisfied: (1) the collision frequency and mode frequency are sufficiently close $\nu \approx \omega_*$ (or more precisely, the MTM dispersion relation produces instability), and (2) the relevant rational surface aligns with the peak in $\omega_*$.  The latter criterion is particularly relevant in pedestals with low magnetic shear and/or where MTMs are stimulated at low toroidal mode number where rational surfaces are more sparse.  In such scenarios, MTMs will be unstable at disparate $n$ numbers, as is often observed in magnetic spectrograms, the present study being representative.  In other scenarios where the resonance occurs at higher $n$ or where magnetic shear is sufficiently high to produce a continuum of MTM instability, MTMs will likely be manifest as a broader band of fluctuations over a range of consecutive $n$ numbers, as is reported in Ref.~\cite{perez_03}.  

\begin{figure}[htb!]
 \centering
 \includegraphics{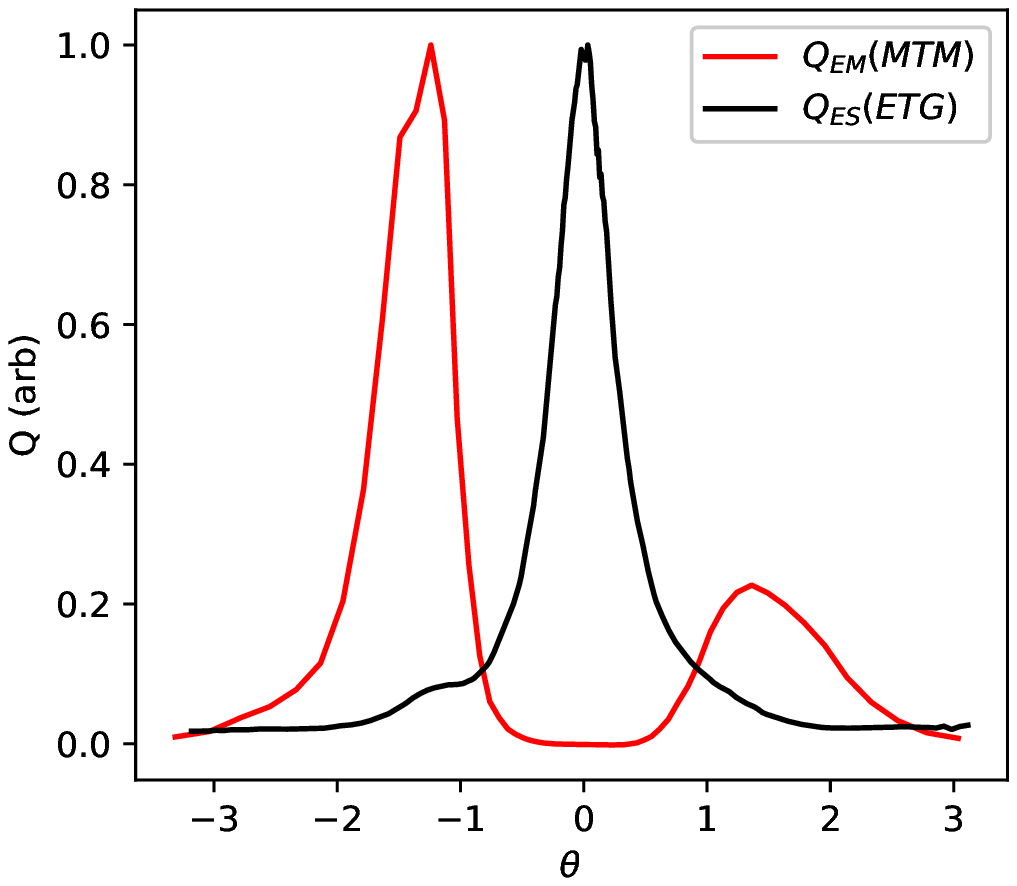}
 \caption{\label{fluxz_profs} Poloidal distribution of heat flux from MTMs (red) and ETG (black).  The MTM simulation is global and limited to low $k_y$, whereas the ETG simulation produces eat flux at very small scales ($k_y \rho_s \approx 80$).  For reference, the ETG simulation is a local simulation at $\rho_{tor}=0.975$ and produces 4 MW of transport. }
\end{figure}

The propensity of MTMs to arise at the peak in the $\omega_*$ profile is likely to set an upper bound on the maximum temperature gradient accessible in a given pedestal scneario.  Once MTMs sufficiently surpass the linear threshold, they form magnetic islands and rapidly relax the background electron temperature profile in a region surrounding the rational surface.  ETG transport is likely to be closely linked to these dynamics as follows.  The flattening of the background temperature profile will simultaneously create localized regions of steep temperature gradients at the boundaries of the islands (see Fig.~\ref{nonlinear}), which will in turn produce fine-scale ETG turbulence in regions of enhanced electron temperature gradients.  Other interaction between MTMs and ETG is also possible, for example, direct nonlinear coupling between scales and/or the interaction of both with background flows and fields.  Direct coupling may be rather unlikely considering the distinct poloidal locations at which each transport mechanism operates, with MTMs favoring the top/bottom of the tokamak and ETG favoring the outboard midplane as shown in Fig.~\ref{fluxz_profs}.  Interaction with lower $n$ toroidal ETG modes~\cite{parisi_20} is also an important consideration.  These fascinating multi-scale interactions are the topic of ongoing study with an initial investigation reported in Ref.~\cite{pueschel_20}.  Earlier work for core-like parameters reported in Ref.~\cite{maeyama_17}.  ETG is also the most likely candidate to account for electron heat transport across the regions of the pedestal not limited by MTMs.    


Other transport channels (impurities, ion temperature, density) are likely mediated by separate transport mechanisms as described in detail in~\cite{kotschenreuther_19,TPT}. 


\section{Discussion and Conclusions} \label{summary}

This paper describes a theoretical and numerical examination of magnetic fluctuations observed in the JET pedestal, unambiguously identifying them as MTMs.  Gyrokinetic simulations quantitatively capture many distinctive features of the experimentally observed fluctuations, including the width, peaks, and toroidal mode numbers associated with two prominent frequency bands observed in magnetic spectrograms.  A major achievement of this work is a clear explanation for the discrete bands observed at disparate toroidal mode numbers $n=4,8$.  This distinctive feature of the fluctuations is due to the alignment of the $n=4,8$ rational surfaces with the peak in the $\omega_*$ profile and the misalignment of the rational surface corresponding to the other $n$ numbers.  Comparisons with a simple MTM dispersion relation were shown, exhibiting good agreement on the collisionality threshold and qualitative agreement on the growth rate dependence on collision frequency.  Although some modes were found to have collisionless tails where MTMs remain unstable, the experimental points lie in the collisional or semi-collisional regimes.  Despite some limitations, nonlinear simulations elucidate many aspects of the experimental observations and offer valuable physical insights.  The MTMs saturate by relaxing the background temperature gradient.  The resulting local reduction brings the frequencies into close agreement with the experimental observations.  We posit that ETG and MTMs closely interact to account for electron heat transport across the pedestal.  ETG likely serves as the main electron heat transport mechanism in regions of the pedestal outside the MTM region and likely serves to reinforce the background temperature gradient as it is relaxed by MTMs.  
This paper further establishes the physical basis for the origin, role, and distinctive characteristics of an important and commonly observed pedestal fluctuation.     

{\em Acknowledgements.--} This research used resources of the National Energy Research Scientific Computing Center, a DOE Office of Science User Facility.  We acknowledge the CINECA award under the ISCRA initiative, for the availability of high performance computing resources and support.  This work was supported by U.S. DOE Contract No. DE-FG02-04ER54742 and U.S. DOE Office of Fusion Energy Sciences Scientific Discovery through Advanced Computing (SciDAC) program under Award Number DE-SC0018429. This work has been carried out within the framework of the EUROfusion Consortium and has received funding from the Euratom research and training programme 2014-2018 and 2019-2020 under grant agreement No 633053. The views and opinions expressed herein do not necessarily reflect those of the European Commission. 

\appendix

\section{Numerical setup of gyrokinetic simulations}
\label{appendix}

This work used the \textsc{Gene} code mostly in its global mode of operation.  Numerical details of the simulations are described in this appendix.  All simulations were electromagnetic and employed a Landau-Boltzmann collision operator~\cite{doerk_dissertation} with collision frequencies defined by the experimental conditions.  All global simulations include background $E \times B$ shear.  The recently-developed block-structured velocity space grids~\cite{jarema_16,jarema_17}  were exploited to reduce the demands on velocity space resolution, accelerating global simulations.    

Global simulations use 320 radial grid points and span the domain $\rho_{tor}=0.94-0.995$.  Dirichlet boundary conditions were enforced at the radial boundaries and transition regions were implemented (10 \% on each side) over which gradients are smoothly set to zero and Krook damping smoothly ramps up to set fluctuations to zero at the boundary.  

In the parallel $z$ direction, $60-64$ grid points were employed.  In parallel velocity $v_{||}$, and magnetic moment $\mu$ coordinates (i.e. squared perpendicular velocity), ($60,24$) grid points were used, respectively, in the absence of block-structured grids.  This could be reduced to ($\sim 36,12$) when block-structured grids were employed.  The parallel domain was from $-\pi$ to $\pi$ (poloidal angle), and the velocity space domains are adapted radially to span approximately $-3$ to $3$ in the parallel velocity coordinate (normalized to $\sqrt{T_e/m_i}$), and $0$ to $9$ (normalized to $T_e/B_0$) in the $\mu$ coordinate.  In the block-structured grids approach, these domains are adapated radially corresponding to the local temperatures.           
Source rates ranging from 0.05 to 0.5 were tested in the nonlinear simulations.  Values above 0.35 generally prevented the relaxation of the background gradients but resulted in numerical instabilities early in the simulations.  Values in the range 0.1-0.2 were used for the simulations shown in Fig.~\ref{nonlinear}, for which the background gradients relaxed and subsequently rebuilt.

Fourth order hyperdiffusion~\cite{pueschel_10} was employed in the parallel, radial, and parallel velocity coordinates.  Coefficients in the radial direction range from 2-10, in the parallel velocity direction 0.2, and in the parallel direction the coefficient was set to the corresponding value for an upwinding scheme.  

\section{References}

\bibliography{my_refs}{}

\begin{thebibliography}{10}

\bibitem{perez_03}
C.~P. Perez, H.~R. Koslowski, T.~C. Hender, P.~Smeulders, A.~Loarte, P.~J.
  Lomas, G.~Saibene, R.~Sartori, M.~Becoulet, T.~Eich, and {others}.
\newblock Washboard modes as {ELM}-related events in {JET}.
\newblock {\em Plasma physics and controlled fusion}, 46(1):61, 2003.

\bibitem{diallo_14}
A.~Diallo, J.~W. Hughes, M.~Greenwald, B.~LaBombard, E.~Davis, S-G. Baek,
  C.~Theiler, P.~Snyder, J.~Canik, J.~Walk, T.~Golfinopoulos, J.~Terry,
  M.~Churchill, A.~Hubbard, M.~Porkolab, L.~Delgado-Aparicio, M.~L. Reinke,
  A.~White, and Alcator C-Mod team.
\newblock Observation of edge instability limiting the pedestal growth in
  tokamak plasmas.
\newblock {\em Phys. Rev. Lett.}, 112:115001, Mar 2014.

\bibitem{diallo_15}
A.~Diallo, R.~J. Groebner, T.~L. Rhodes, D.~J. Battaglia, D.~R. Smith, T.~H.
  Osborne, J.~M. Canik, W.~Guttenfelder, and P.~B. Snyder.
\newblock Correlations between quasi-coherent fluctuations and the pedestal
  evolution during the inter-edge localized modes phase on diii-d.
\newblock {\em Physics of Plasmas}, 22(5):056111, 2015.

\bibitem{laggner_16}
F.~M. Laggner, E.~Wolfrum, M.~Cavedon, F.~Mink, E.~Viezzer, M.~G. Dunne,
  P.~Manz, H.~Doerk, G.~Birkenmeier, R.~Fischer, {S Fietz}, M.~Maraschek,
  M.~Willensdorfer, F.~Aumayr, the EUROfusion~MST1 Team, and the ASDEX~Upgrade
  Team.
\newblock High frequency magnetic fluctuations correlated with the inter-{ELM}
  pedestal evolution in {ASDEX} {Upgrade}.
\newblock {\em Plasma Physics and Controlled Fusion}, 58(6):065005, 2016.

\bibitem{cavedon_19}
M.~Cavedon, R.~Dux, T.~Pütterich, E.~Viezzer, E.~Wolfrum, M.~Dunne, E.~Fable,
  R.~Fischer, G.F. Harrer, F.M. Laggner, A.F. Mink, U.~Plank, U.~Stroth,
  M.~Willensdorfer, and Asdex Upgrade~Team.
\newblock On the ion and electron temperature recovery after the {ELM}-crash at
  {ASDEX} upgrade.
\newblock {\em Nuclear Materials and Energy}, 18:275--280, January 2019.

\bibitem{chen_19}
J.~Chen.
\newblock Internal measurement of pedestal-localized broadband magnetic
  fluctuations in elmy h-mode plasmas in diii-d.
\newblock {\em Bulleting of the American Physical Society}.

\bibitem{laggner_19}
F.M. Laggner, A.~Diallo, M.~Cavedon, and E.~Kolemen.
\newblock Inter-elm pedestal localized fluctuations in tokamaks: Summary of
  multi-machine observations.
\newblock {\em Nuclear Materials and Energy}, 19:479 -- 486, 2019.

\bibitem{jenko_00b}
F.~Jenko, W.~Dorland, M.~Kotschenreuther, and B.N. Rogers.
\newblock Electron temperature gradient driven turbulence.
\newblock {\em Phys. Plasmas}, 7:1904, 2000.

\bibitem{goerler_11}
T.~G\"orler, X.~Lapillonne, S.~Brunner, T.~Dannert, F.~Jenko, F.~Merz, and
  D.~Told.
\newblock The global version of the gyrokinetic turbulence code {GENE}.
\newblock {\em Journal of Computational Physics}, 230(18):7053--7071, August
  2011.

\bibitem{HDW}
R.~D. Hazeltine, D.~Dobrott, and T.~S. Wang.
\newblock Kinetic theory of tearing instability.
\newblock {\em Phys. Fluids}, 18(12):1778, 1975.

\bibitem{mahajan_79}
Swadesh~M. Mahajan, R.~D. Hazeltine, H.~R. Strauss, and David~W. Ross.
\newblock Unified theory of tearing modes.
\newblock {\em The Physics of Fluids}, 22(11):2147--2157, 1979.

\bibitem{gladd_80}
N.~T. Gladd, J.~F. Drake, C.~L. Chang, and C.~S. Liu.
\newblock Electron temperature gradient driven microtearing mode.
\newblock {\em Phys. Fluids}, 23(6):1182, 1980.

\bibitem{carmody_13}
D.~Carmody, M.~J. Pueschel, and P.~W. Terry.
\newblock Gyrokinetic studies of microinstabilities in the reversed field
  pinch.
\newblock {\em Physics of Plasmas}, 20(5):052110, 2013.

\bibitem{predebon_13}
I.~Predebon and F.~Sattin.
\newblock On the linear stability of collisionless microtearing modes.
\newblock {\em Physics of Plasmas}, 20(4):040701--040701--4, April 2013.

\bibitem{swamy_14}
Aditya~K. Swamy, R.~Ganesh, J.~Chowdhury, S.~Brunner, J.~Vaclavik, and
  L.~Villard.
\newblock Global gyrokinetic stability of collisionless microtearing modes in
  large aspect ratio tokamaks.
\newblock {\em Physics of Plasmas}, 21(8):082513, 2014.

\bibitem{hamed_18}
M.~Hamed, M.~Muraglia, Y.~Camenen, and X.~Garbet.
\newblock Stability of a slab collisional microtearing mode.
\newblock {\em Contributions to Plasma Physics}, 58(6-8):529--533.

\bibitem{doerk_11}
H.~Doerk, F.~Jenko, M.~J. Pueschel, and D.~R. Hatch.
\newblock Gyrokinetic {Microtearing} {Turbulence}.
\newblock {\em Physical Review Letters}, 106(15), April 2011.

\bibitem{doerk_12}
H.~Doerk, F.~Jenko, T.~Görler, D.~Told, M.~J. Pueschel, and D.~R. Hatch.
\newblock Gyrokinetic prediction of microtearing turbulence in standard
  tokamaks.
\newblock {\em Physics of Plasmas}, 19(5):055907, 2012.

\bibitem{guttenfelder_11}
W.~Guttenfelder, J.~Candy, S.~M. Kaye, W.~M. Nevins, E.~Wang, R.~E. Bell, G.~W.
  Hammett, B.~P. LeBlanc, D.~R. Mikkelsen, and H.~Yuh.
\newblock Electromagnetic {Transport} from {Microtearing} {Mode} {Turbulence}.
\newblock {\em Phys. Rev. Lett.}, 106(15):155004, April 2011.

\bibitem{applegate_07}
D~J Applegate, C~M Roach, J~W Connor, S~C Cowley, W~Dorland, R~J Hastie, and
  N~Joiner.
\newblock Micro-tearing modes in the mega ampere spherical tokamak.
\newblock {\em Plasma Physics and Controlled Fusion}, 49(8):1113, 2007.

\bibitem{chowdhury_16}
J.~Chowdhury, Yang Chen, Weigang Wan, Scott~E. Parker, W.~Guttenfelder, and
  J.~M. Canik.
\newblock Particle-in-cell δf gyrokinetic simulations of the microtearing
  mode.
\newblock {\em Physics of Plasmas (1994-present)}, 23(1):012513, January 2016.

\bibitem{dickinson_12}
D.~Dickinson, C.~M. Roach, S.~Saarelma, R.~Scannell, A.~Kirk, and H.~R. Wilson.
\newblock Kinetic instabilities that limit $\ensuremath{\beta}$ in the edge of
  a tokamak plasma: A picture of an $h$-mode pedestal.
\newblock {\em Phys. Rev. Lett.}, 108:135002, Mar 2012.

\bibitem{canik_13}
J.M. Canik, W.~Guttenfelder, R.~Maingi, T.H. Osborne, S.~Kubota, Y.~Ren, R.E.
  Bell, H.W. Kugel, B.P. LeBlanc, and V.A. Souhkanovskii.
\newblock Edge microstability of nstx plasmas without and with lithium-coated
  plasma-facing components.
\newblock {\em Nuclear Fusion}, 53(11):113016, 2013.

\bibitem{jian_19}
X.~Jian, C.~Holland, J.~Candy, E.~Belli, V.~Chan, A. M. Garofalo, and
  S.~Ding.
\newblock Role of {Microtearing} {Turbulence} in {DIII}-{D} {High} {Bootstrap}
  {Current} {Fraction} {Plasmas}.
\newblock {\em Phys. Rev. Lett.}, 123(22):225002, November 2019.

\bibitem{hatch_16}
D.~R. Hatch, M.~Kotschenreuther, S.~Mahajan, P.~Valanju, F.~Jenko, D.~Told,
  T.~G\"orler, and S.~Saarelma.
\newblock Microtearing turbulence limiting the {JET}-{ILW} pedestal.
\newblock {\em Nuclear Fusion}, 56(10):104003, 2016.

\bibitem{kotschenreuther_19}
M.~Kotschenreuther, X.~Liu, D.R. Hatch, S.~Mahajan, L.~Zheng, A.~Diallo,
  R.~Groebner, {the DIII-D TEAM}, J.C. Hillesheim, C.F. Maggi, C.~Giroud,
  F.~Koechl, V.~Parail, S.~Saarelma, E.~Solano, A.~Chankin, and {JET
  Contributors}.
\newblock Gyrokinetic analysis and simulation of pedestals to identify the
  culprits for energy losses using ‘fingerprints’.
\newblock {\em Nucl. Fusion}, 59(9):096001, September 2019.

\bibitem{chen_16}
J.~Chen, W.~X. Ding, D.~L. Brower, D.~Finkenthal, C.~Muscatello, D.~Taussig,
  and R.~Boivin.
\newblock Faraday-effect polarimeter diagnostic for internal magnetic field
  fluctuation measurements in diii-d.
\newblock {\em Review of Scientific Instruments}, 87(11):11E108, 2016.

\bibitem{hatch_19}
D.R. Hatch, M.~Kotschenreuther, S.M. Mahajan, G.~Merlo, A.R. Field, C.~Giroud,
  J.C. Hillesheim, C.F. Maggi, C.~Perez~von Thun, C.M. Roach, S.~Saarelma, and
  {JET Contributors}.
\newblock Direct gyrokinetic comparison of pedestal transport in {JET} with
  carbon and {ITER}-like walls.
\newblock {\em Nucl. Fusion}, 59(8):086056, August 2019.

\bibitem{landreman_12}
Matt Landreman and Darin~R Ernst.
\newblock Local and global fokker–planck neoclassical calculations showing
  flow and bootstrap current modification in a pedestal.
\newblock {\em Plasma Physics and Controlled Fusion}, 54(11):115006, 2012.

\bibitem{viezzer_13}
E.~Viezzer, T.~Pütterich, G.D. Conway, R.~Dux, T.~Happel, J.C. Fuchs, R.M.
  McDermott, F.~Ryter, B.~Sieglin, W.~Suttrop, M.~Willensdorfer, E.~Wolfrum,
  and the ASDEX Upgrade~Team.
\newblock High-accuracy characterization of the edge radial electric field at
  asdex upgrade.
\newblock {\em Nuclear Fusion}, 53(5):053005, 2013.

\bibitem{dickinson_13}
D~Dickinson, C~M Roach, S~Saarelma, R~Scannell, A~Kirk, and H~R Wilson.
\newblock Microtearing modes at the top of the pedestal.
\newblock {\em Plasma Physics and Controlled Fusion}, 55(7):074006, 2013.

\bibitem{saarelma_13}
S.~Saarelma, M.N.A. Beurskens, D.~Dickinson, L.~Frassinetti, M.J. Leyland, C.M.
  Roach, and EFDA-JET Contributors.
\newblock Mhd and gyro-kinetic stability of jet pedestals.
\newblock {\em Nuclear Fusion}, 53(12):123012, 2013.

\bibitem{hastie_81}
R.J. Hastie and K.W. Hesketh.
\newblock Kinetic modifications to the {MHD} ballooning mode.
\newblock {\em Nuclear Fusion}, 21(6):651--656, jun 1981.

\bibitem{cheng_82}
C.~Z. Cheng.
\newblock Kinetic theory of collisionless ballooning modes.
\newblock {\em The Physics of Fluids}, 25(6):1020--1026, 1982.

\bibitem{kotschenreuther_86}
M.~Kotschenreuther.
\newblock Compressibility effects on ideal and kinetic ballooning modes and
  elimination of finite larmor radius stabilization.
\newblock {\em The Physics of Fluids}, 29(9):2898--2913, 1986.

\bibitem{parra_15}
Felix~I Parra and Michael Barnes.
\newblock Equivalence of two different approaches to global delta f gyrokinetic
  simulations.
\newblock {\em Plasma Physics and Controlled Fusion}, 57(5):054003, apr 2015.

\bibitem{abdoul_15}
P~A Abdoul, D~Dickinson, C~M Roach, and H~R Wilson.
\newblock Using a local gyrokinetic code to study global ion temperature
  gradient modes in tokamaks.
\newblock {\em Plasma Physics and Controlled Fusion}, 57(6):065004, apr 2015.

\bibitem{field_20}
A~R Field, C~D Challis, J~M Fontdecaba, L~Frassinetti, L~Horvath, Hyun-Tae Kim,
  C~Maggi, C~M Roach, S~Saarelma, M~Sertoli, and G~Szepeisi and.
\newblock The dependence of exhaust power components on edge gradients in
  {JET}-c and {JET}-{ILW} h-mode plasmas.
\newblock {\em Plasma Physics and Controlled Fusion}, 62(5):055010, mar 2020.

\bibitem{larakers_20}
J.~L. Larakers, R.~D. Hazeltine, and S.~M. Mahajan.
\newblock A comprehensive conductivity model for drift and micro-tearing modes.
\newblock {\em Physics of Plasmas}, 27(6):062503, 2020.

\bibitem{hassan_20}
E.~Hassan, D.~R. Hatch, M.~Halfmoon, G~Merlo, A.~O. Nelson, M.~Kotschenreuther,
  S.~Mahajan, R.~J. Groebner, and A.~Diallo.
\newblock Clear signature of micro-tearing modes in the diii-d pedestal.
\newblock {\em in preparation}.

\bibitem{pueschel_20}
M.~J. Pueschel and \textit{et al.}
\newblock Multi-scale interactions of microtearing turbulence in the tokamak
  pedestal.
\newblock {\em in preparation}.

\bibitem{maeyama_17}
S.~Maeyama, T.-H. Watanabe, and A.~Ishizawa.
\newblock Suppression of {Ion}-{Scale} {Microtearing} {Modes} by
  {Electron}-{Scale} {Turbulence} via {Cross}-{Scale} {Nonlinear}
  {Interactions} in {Tokamak} {Plasmas}.
\newblock {\em Physical Review Letters}, 119(19), November 2017.

\bibitem{wolfrum_09}
E.~Wolfrum, A.~Burckhart, R.~Fischer, N.~Hicks, C.~Konz, B.~Kurzan, B.~Langer,
  T.~Pütterich, H.~Zohm, and the ASDEX~Upgrade Team.
\newblock Investigation of inter-{ELM} pedestal profiles in {ASDEX} {Upgrade}.
\newblock {\em Plasma Phys. Control. Fusion}, 51(12):124057, December 2009.

\bibitem{maggi_17}
C.F. Maggi, L.~Frassinetti, L.~Horvath, A.~Lunniss, S.~Saarelma, H.~Wilson,
  J.~Flanagan, M.~Leyland, I.~Lupelli, S.~Pamela, H.~Urano, L.~Garzotti,
  E.~Lerche, I.~Nunes, F.~Rimini, and JET Contributors.
\newblock Studies of the pedestal structure and inter-elm pedestal evolution in
  jet with the iter-like wall.
\newblock {\em Nuclear Fusion}, 57(11):116012, 2017.

\bibitem{TPT}
D.~R. Hatch and \textit{et al.}
\newblock Final report for the fy19 fes theory performance target.
\newblock Technical report, 2019.

\bibitem{parisi_20}
Jason~F. Parisi, Felix~I. Parra, Colin~M. Roach, Carine Giroud, William
  Dorland, David~R. Hatch, Michael Barnes, Jon~C. Hillesheim, Nobuyuki Aiba,
  Justin Ball, Plamen~G. Ivanov, and JET Contributors.
\newblock Toroidal and slab etg instability dominance in the linear spectrum of
  jet-ilw pedestals, 2020.

\bibitem{doerk_dissertation}
H.~Doerk.
\newblock Gyrokinetic simulation of microtearing turbulence.
\newblock {\em Dissertation, University of Ulm}, 2013.

\bibitem{jarema_16}
D.~Jarema, H.~J. Bungartz, T.~Görler, F.~Jenko, T.~Neckel, and D.~Told.
\newblock Block-structured grids for {Eulerian} gyrokinetic simulations.
\newblock {\em Computer Physics Communications}, 198:105--117, January 2016.

\bibitem{jarema_17}
D.~Jarema, H.~J. Bungartz, T.~Görler, F.~Jenko, T.~Neckel, and D.~Told.
\newblock Block-structured grids in full velocity space for {Eulerian}
  gyrokinetic simulations.
\newblock {\em Computer Physics Communications}, 215:49 -- 62, 2017.

\bibitem{pueschel_10}
M.J. Pueschel, T.~Dannert, and F.~Jenko.
\newblock On the role of numerical dissipation in gyrokinetic {Vlasov}
  simulations of plasma microturbulence.
\newblock {\em Computer Physics Communications}, 181(8):1428--1437, August
  2010.

\end{thebibliography}
\bibliographystyle{unsrt}

\end{document}